\documentclass[10pt,aps,prd,twocolumn,nofootinbib,preprintnumbers,floatfix,superscriptaddress,amsmath,amssymb,altaffilletter]{revtex4-2}
\usepackage[colorlinks=true,pdfstartview=FitV,breaklinks=true]{hyperref}

\usepackage{subfigure}
\usepackage{times}
\usepackage{graphicx}
\usepackage{dcolumn}
\usepackage{bm}
\usepackage{tensor}
\usepackage{slashed}
\usepackage{bm}
\usepackage{multirow}
\usepackage{soul}
\usepackage{amsmath}
\usepackage{fontawesome}
\usepackage{mathrsfs}
\usepackage{amssymb}
\usepackage{yfonts}
\usepackage{color}
\usepackage{xspace}
\usepackage{url}
\usepackage{verbatim}
\usepackage{mathtools}
\usepackage{upgreek}
\usepackage{units}
\usepackage{siunitx}
\usepackage{amstext}
\usepackage{booktabs}
\usepackage{tabulary}
\usepackage{tabularx}
\usepackage{etoolbox}
\usepackage[version=4]{mhchem}
\usepackage[utf8]{inputenc}
\usepackage[dvipsnames,table]{xcolor}
\newcolumntype{Y}{>{\centering\arraybackslash}X}
\hypersetup{urlcolor=BlueViolet,
	    citecolor=Plum,
	    linkcolor=NavyBlue}
\usepackage[official]{eurosym}
\definecolor{lightgray}{rgb}{0.9,0.9,0.9}	    
\definecolor{green}{rgb}{0,0.5,0}
\definecolor{red}{rgb}{1,0,0}
\definecolor{blue}{rgb}{0,0,0.5}

 \newcommand{\rhodm}{\rho_\mathrm{DM}}
 
 \newcommand{\vbf}{\mathbf{v}}

\begin{document}

\preprint{ \includegraphics[width=0.37\textwidth]{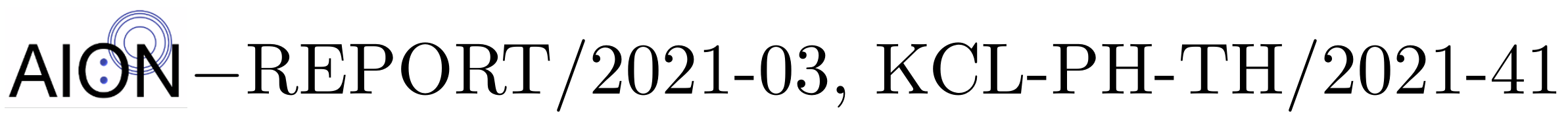}  }

\title{Refined ultralight scalar dark matter searches with\\ compact atom gradiometers}

\author{Leonardo Badurina}
\email{leonardo.badurina@kcl.ac.uk}
\affiliation{Theoretical Particle Physics and Cosmology Group, Department of Physics, King's College London, Strand, London, WC2R 2LS, UK}

\author{Diego Blas}
\affiliation{Grup  de  F\'isica  Te\`orica,  Departament  de  F\'isica,  Universitat  Aut\`onoma  de  Barcelona,  08193  Bellaterra, Spain,}
\affiliation{Institut de Fisica d’Altes Energies (IFAE), The Barcelona Institute of Science and Technology,\\ Campus UAB, 08193 Bellaterra (Barcelona), Spain}
\affiliation{Theoretical Particle Physics and Cosmology Group, Department of Physics, King's College London, Strand, London, WC2R 2LS, UK}

\author{Christopher McCabe}
\affiliation{Theoretical Particle Physics and Cosmology Group, Department of Physics, King's College London, Strand, London, WC2R 2LS, UK}
             
\begin{abstract}
Atom interferometry is a powerful experimental technique that can be employed to search for the oscillation of atomic transition energies induced by ultra-light scalar dark matter (ULDM).
Previous studies have focused on the sensitivity to ULDM of km-length atom gradiometers, where atom interferometers are located at the ends of very long baselines.
In this work, we generalise the treatment of the time-dependent signal induced by a linearly-coupled scalar ULDM candidate for vertical atom gradiometers of any length and find correction factors that especially impact the ULDM signal in short-baseline gradiometer configurations. 
Using these results, we refine the sensitivity estimates in the limit where shot noise dominates for AION-10, a compact \SI{10}{m} gradiometer that will be operated in Oxford, and discuss optimal experimental parameters that enhance the reach of searches for linearly-coupled scalar ULDM. After comparing the reach of devices operating in broadband and resonant modes, we show that well-designed compact atom gradiometers are able to explore regions of dark matter parameter space that are not yet constrained.
\end{abstract}

\maketitle

\section{Introduction}

Despite overwhelming astrophysical and cosmological evidence for its existence through gravitationally induced effects, dark matter (DM) has yet to be understood at a fundamental level~\cite{Bertone:2004pz}. Indeed, the fundamental constituents of DM and their interactions are completely unknown, and the DM mass could range from particles as light as 10\textsuperscript{-22} eV to asteroid-mass primordial black holes~\cite{Battaglieri:2017aum, Green:2020jor}. To unravel the constituents of DM, direct searches for DM particles via non-gravitational interactions have been among the highest priorities in particle physics, astrophysics, and cosmology~\cite{APPEC}. 

Because of null results in direct searches for weak mass-scale particles in colliders and direct detection searches (see, e.g., the constraints from the xenon-based experiments~\cite{Aprile:2018dbl,PandaX:2021osp}), increasing experimental and theoretical interest has shifted to ultra-light bosons with a sub-eV mass. 
Owing to their large occupation number in the Milky Way, such ultra-light dark matter (ULDM) candidates are expected to behave as coherent oscillating waves, and thus would lead to new phenomenological signatures~\cite{Ferreira:2020fam}. 
Prominent among these ultra-light bosons are many well-motivated DM candidates~\cite{Jaeckel:2010ni, Hui:2021tkt}, such as dilatons, moduli, and the relaxion~\cite{Graham:2015cka}, as well as the pseudoscalar QCD axion and axion-like-particles (ALPs)~\cite{Marsh:2015xka}, and vector bosons~\cite{Nelson:2011sf,Graham:2015ifn}.

Multiple phenomenological `portals' have been proposed to test the wide range of well-motivated theoretical models over as wide a range of masses as possible. 
In this paper we focus on the interactions between ultra-light {\it scalar} DM with electrons or photons, which can lead to oscillations in fundamental constants (see e.g.~\cite{Arvanitaki:2014faa, Stadnik:2014tta,Stadnik:2015kia}). These DM-induced oscillations can in turn alter atomic transition energies, which can be searched for with atom interferometer experiments~\cite{Geraci:2016fva,Arvanitaki:2016fyj}. 

An atom interferometer can be succinctly described as an experiment that compares the  
phase between coherent spatially delocalised quantum superpositions of atom clouds~\cite{Abend:2020djo}.  
Two spatially separated atom interferometers that are referenced by the same laser pulse via single-photon atomic transitions can be operated as a gradiometer~\cite{Graham:2012sy}. 
The distinctive signal of a coherent oscillating DM field can be tested with a gradiometer by finding the difference between the phases measured by two atom interferometers~\cite{Arvanitaki:2016fyj}. 
A key advantage of the gradiometer experimental set-up over a lone single-photon atom interferometer relies on the possibility of effectively attenuating laser noise. By operating the same laser pulse between both interferometers, the interferometers' laser noise cancels in a differential measurement~\cite{Graham:2012sy}.

A number of single-photon atom gradiometers have been proposed or are currently under construction, including AION-10~\cite{Badurina:2019hst}, MAGIS-100~\cite{Coleman:2018ozp,Abe:2021ksx}, MIGA~\cite{Canuel:2017rrp}, ELGAR~\cite{Canuel:2019abg}, ZAIGA~\cite{ZAIGA}, and  AEDGE~\cite{Bertoldi:2019tck}.
Many of these experiments are envisaged as a test-bed for technologies that could later be employed in much larger km-scale interferometers, which is considered to be the scale required in order to operate as a mid-band gravitational wave detector~\cite{ Dimopoulos:2007cj,Graham:2016plp,Graham:2017lmg, Ellis:2020lxl}. For example, within the AION project, AION-10 is a \SI{10}{m} instrument that will be operated in Oxford and there is a roadmap in place to later develop \SI{100}{m} and km versions~\cite{Badurina:2019hst}. 

While the DM-induced phase difference has been calculated for km-scale gradiometers, a careful treatment accounting for the DM-induced phase accumulated by the atoms along each path segment has not been provided for more compact designs, such as AION-10. Remarkably this compact set-up could be sensitive to a currently unconstrained region of DM models, making this analysis even more relevant. The purpose of this paper is therefore to refine the previous calculations in~\cite{Arvanitaki:2016fyj,Badurina:2019hst}, and to provide updated projections of the AION-10 sensitivity to linear ultra-light scalar DM--electron and photon interactions. 

This paper is structured as follows. 
In section~\ref{sec:theomod} we define the ULDM model and the parameters that can be constrained experimentally. 
In section~\ref{sec:ULDMsig}, we present the refined DM-phase calculation for short-baseline gradiometers.
In section~\ref{sec:sensitivityscaling}, we discuss how the sensitivity of a compact gradiometer depends on tuneable experimental parameters.
Then in section~\ref{sec:optAION}, we utilise the short-baseline results to provide updated sensitivity estimates for the AION-10 experiment assuming aspirational parameters for the atom numbers and other
performance factors. In particular, we work in the limit where shot noise dominates and do not include the effects of technical noise. Finally, we present our conclusions in section~\ref{sec:conclusions}.
Two appendices provide further details and derivations of calculations to support the results in section~\ref{sec:ULDMsig}.

\section{\label{sec:theomod} Ultra-light Scalar Dark Matter}

Ultra-light dark matter (ULDM) can be modeled as a temporally and spatially oscillating non-relativistic classical field, which follows from its high occupation number and its small mean velocity and velocity dispersion (characteristic of DM in the Milky Way)~\cite{Hui:2021tkt}. 

The velocity $\vbf$ of the virialized DM has a typical magnitude of $10^{-3}$ in natural units and follows some astrophysical velocity distribution $f(\vbf)$. In this work, the precise form of the velocity distribution does not need to be specified; we only use the fact that the spread of DM velocities $\sigma_v \approx160~\mathrm{km}\,\mathrm{s}^{-1}\approx 5\times 10^{-4}$ (in natural units)~\cite{Evans:2018bqy} and that the field is non-relativistic. Temporally, the field will be coherent over a coherence time $\tau_c = 2\pi /m_{\phi} \sigma_v^2$, while spatially it will also be coherent over a coherence length $\lambda_c = 2\pi/m_{\phi} \sigma_v$~\cite{Centers:2019dyn}.

Within a coherence length and time of the field, the light scalar~$\phi$ is modelled by the classical non-relativistic solution of the equations of motion for a Lagrangian describing a free massive scalar particle
\begin{equation}
\phi(t, \mathbf{x}) \simeq \frac{\sqrt{2\rhodm}}{m_\phi} \cos{\left ( \omega_{\phi} t - \textbf{k}_{\phi}\cdot \textbf{x}+\theta \right )} \, ,
\label{eq:phiosc}
\end{equation}
where $\omega_{\phi} = m_\phi(1+|\textbf{v}|^2/2) + \mathcal{O}(m_\phi^2 |\textbf{v}|^3)$, $\textbf{k}_{\phi} = m_\phi\textbf{v}$, and~$\theta$ is an unknown phase that is assumed to be random.\footnote{We use natural units with $\hbar=c=1$.} We see that the field is characterized by its temporal angular frequency, which is set by the dark matter mass $m_\phi$ up to small kinetic energy corrections. We have assumed that the field saturates the entire local DM density $\rhodm$ so that the amplitude is equal to $\sqrt{2\rhodm}/m_{\phi}$. In our numerical results, we use the canonical value $\rhodm=\SI{0.3}{\mathrm{GeV}\,\mathrm{cm}^{-3}}$~\cite{Read:2014qva}.

In this paper, we will only consider the leading linear-couplings of \(\phi \) with the Standard Model (SM) fields. 
We will consider the effective Lagrangian at an energy scale of $\sim 1$ GeV describing an ultra-light scalar DM particle coupled linearly to SM fields~\cite{Kaplan:2000hh}. Explicitly, the Lagrangian that we  consider~is
\begin{equation}
\mathcal{L} \supset \mathcal{L}_{\mathrm{SM}}+\mathcal{L}_{\phi} \, ,
\end{equation}
in which the pertinent SM Lagrangian terms for electrons and photons are
\begin{equation}
\mathcal{L}_{\mathrm{SM}} \supset -\frac{1}{4 e^{2}} F_{\mu \nu} F^{\mu \nu}-m_{e} \overline{\psi}_e\psi_e \, ,
\end{equation}
where $e$ is the electric charge and $m_e$ is the electron mass, $F_{\mu \nu}$ is the Maxwell field-strength tensor, and the linear interactions between the ultra-light scalar DM and the photon and electron fields~are
\begin{equation}
\begin{aligned}
\mathcal{L}_{\phi} & \supset \phi(t,\mathbf{x}) \sqrt{4 \pi G_{N}} \left[\frac{d_{e}}{4 e^{2}} F_{\mu \nu} F^{\mu \nu} -d_{m_{e}} m_{e} \overline{\psi}_e\psi_e \right].
\end{aligned}
\label{DM-SM lin couplings}
\end{equation}
Following the convention in Refs.~\cite{Damour:2010rm,Damour:2010rp}, we have parameterised the $d_{e}$ and $d_{m_{e}}$ couplings  relative to Newton's gravitational constant~$G_N$.

The addition of these interactions to the SM Lagrangian results in a time-varying electron mass and fine structure constant~\cite{Damour:2010rp,Stadnik:2014tta}, 
\begin{align}
m_{e}(t, \mathbf{x}) &= m_{e}\left[1+d_{m_{e}}\sqrt{4 \pi G_{N}} \phi(t, \mathbf{x})\right] \\
\alpha(t, \mathbf{x}) & \approx \alpha\left[1+d_{e} \sqrt{4 \pi G_{N}} \phi(t, \mathbf{x})\right] \, .
\end{align}
It is this time-dependence that will give rise to the observable scalar DM signal in an atom gradiometer. 

\section{ULDM Signal in Atom Gradiometers \label{sec:ULDMsig}}

The properties of atoms largely depend on the electron mass $m_e$ and fine structure constant $\alpha$.
For example, for optical electronic transitions, the transition frequency $\omega_A$ scales as $\omega_A \propto m_e \alpha^{2+\xi_A}$ to a high degree~\cite{Arvanitaki:2014faa}, where $\xi_A$ is a calculable parameter governing the electronic transition. 
As a result of the DM-induced oscillations in $m_e$ and $\alpha$, $\omega_A$ will oscillate~as 
\begin{equation}
\omega_A(t, \mathbf{x}) \simeq \omega_A + \Delta \omega_A (t, \mathbf{x}) \, , 
\label{eq: oscillating energy 1}
\end{equation} 
where 
\begin{equation}
 \Delta \omega_A (t, \mathbf{x}) =  \omega_A \sqrt{4\pi G_N} \left [ d_{m_e} + (2+\xi_A) d_e \right] \phi (t, \mathbf{x}) \, .
 \label{eq: oscillating energy 2}
\end{equation}
For our later calculations, we will focus on the AION-10 setup, which will operate on the $\mathrm{5s^2\, ^1S_0 \leftrightarrow 5s 5p \,^3P_0}$ clock-transition in $^{87}\mathrm{Sr}$. For this transition, we have $\omega_A=2.697\times 10^{15}~\mathrm{rad}/\mathrm{s}$ (corresponding to a wavelength $\lambda_A\approx698$~nm)~\cite{Hachisu:17} and $\xi_A\approx0.06$~\cite{Angstmann:2004zz}.

The total phase difference $\Phi$ between two paths of an atom interferometer can be written as the sum of
three contributions (see e.g.,~\cite{Dimopoulos:2008hx}), 
\begin{equation}
\Phi = \Phi_{\mathrm{propagation}} +  \Phi_{\mathrm{laser}}+ \Phi_{\mathrm{separation}}\, .
\end{equation}
The propagation phase, $\Phi_{\mathrm{propagation}}$, arises from the free-fall evolution of the atom's external (e.g.\ linear momentum) and internal (e.g.\ energy levels) degrees of freedom between light pulses; the laser phase, $\Phi_{\mathrm{laser}}$, originates from atom-laser interactions used to manipulate the atom's wavefunction at each beamsplitter and mirror sequence; and the separation phase, $\Phi_{\mathrm{separation}}$, accounts for the wave packet separation of the two `arms' of the interferometer at the final beamsplitter pulse.

ULDM-induced oscillations of fundamental constants would be observable in the evolution of the internal degrees of freedom of atoms and so contribute to the propagation phase. Since the electronic transition energy would receive a spacetime-dependent correction $\Delta \omega_A (t, \mathbf{x})$, an ULDM-induced signal will be accumulated in the propagation phase of the excited state ($\mathrm{5s 5p\,^3P_0}$) relative to that of the ground state ($\mathrm{5s^2\,^1S_0}$) (see Appendix~\ref{A2} for further details).
Explicitly, for each path segment in which the atom is in the excited state, the leading order ULDM-induced phase contribution is~\cite{Arvanitaki:2016fyj}
\begin{equation} 
\Phi_{s_1}^{s_2} \equiv \int_{s_1}^{s_2} {\Delta \omega_A(t, \mathbf{x})\, ds} \,,
\label{integral}
\end{equation}
where the integral is taken along the geodesic of the atom between spacetime points $s_1$ and $s_2$. These points are found by intersecting the atom's and laser's geodesics. Non-classical corrections to the atom geodesics due to the finite size of the atomic wavefunctions are subleading~\cite{Dimopoulos:2008hx} and so will be ignored in our analysis. In addition, effects such as magnetic fields, stray light or blackbody radiation could also lead to corrections to the atom geodesics. Estimates in the context of MAGIS-100 have shown that the contribution to the phase shift from these sources can be controlled such that they are negligible~\cite{Abe:2021ksx}. We expect similar conclusions to apply to a more compact gradiometer such as AION-10 but a detailed study is beyond the scope of this work.

The spacetime on Earth can be effectively modelled using the Schwarzschild metric
\begin{equation}
d s^{2} = \left(1+2 \Phi_{\mathrm{E}}(r)\right) d t^{2} -\frac{1}{1+2 \Phi_{\mathrm{E}}(r)} d r^{2}-r^{2} d \Omega^{2} \, ,
\label{metric}
\end{equation}
where $\Phi_{\mathrm{E}}(r)$ is the Earth's gravitational field at radial position~$r$. Using the metric Eq.~\eqref{metric}, a light pulse that leaves one of the laser sources at position $r_0$ and time $t_0$ will arrive at the atom's at position $r$ at coordinate time~$t(r,r_0,t_0)$. By solving the geodesic equations to third order in the affine parameter~$\lambda$ of the laser's null curve, inverting the laser's radial geodesic~$r(\lambda)$ to determine~$\lambda(r)$ to $\mathcal{O}(r^3)$, and substituting $\lambda(r)$ self-consistently into $t(\lambda)$, we obtain
\begin{equation}
t(r,r_0,t_0)= t_0+\frac{r-r_0}{1+2\Phi_{\mathrm{E}}(r_0)} +\frac{(r-r_0)^2\partial_r \Phi_{\mathrm{E}}(r_0)}{(1+2\Phi_{\mathrm{E}}(r_0))^2} \, .
\label{time}
\end{equation}
On the Earth's surface $\Phi_{\mathrm{E}}(r_{\mathrm{Earth}}) \sim 10^{-8}$, and for all AION proposals, from AION-10 to AION-km, $r-r_0 \lesssim 10^{-5}$~\!s (in natural units). Thus, in subsequent calculations we will make use of the approximation $t(r,r_0,t_0) \approx t_0+r-r_0$.

Furthermore, we will consider interferometric sequences with excited state paths of spatial length $|\Delta \textbf{r}_e|$ much smaller than $\Delta t_e$, which is the duration of these same paths; 
in this case, we may approximate the atom as being stationary when in the excited state, such that the line element $ds$ can be expressed as $\sqrt{g_{00}(r)}dt$.
For the metric in Eq.~\eqref{metric}, $g_{00}(r)=1+2\Phi_{\mathrm{E}}(r)\approx1$, such that we may further use the approximation $ds \approx dt$. Given that $|\textbf{k}_\phi|$ is suppressed by a factor of 10\textsuperscript{-3} relative  to  the  angular  frequency $\omega$,
we may neglect the spatial dependence of $\phi$, and thus of $\Delta \omega_A$, and rewrite Eq.~\eqref{integral} as 
\begin{align} 
\Phi_{t_1}^{t_2} &\equiv \int_{t_1}^{t_2} {\Delta \omega_A(t) dt} \label{integral1} \\
&= \overline{\Delta \omega_A}  \int_{t_1}^{t_2}  \cos (\omega_{\phi} t + \theta)dt\;,
\end{align}
where we have defined
\begin{equation}\label{signal amplitude 2}
    \overline{\Delta \omega_A} = \omega_A \sqrt{4\pi G_N} \left [ d_{m_e} + (2+\xi_A) d_e \right] \frac{\sqrt{2 \rho_\mathrm{DM}}}{m_\phi}\,,
\end{equation}
as the magnitude of $\Delta \omega_A(t,\mathbf{x})$.

\begin{figure}[t]
\centering
    \includegraphics[width=.5\textwidth]{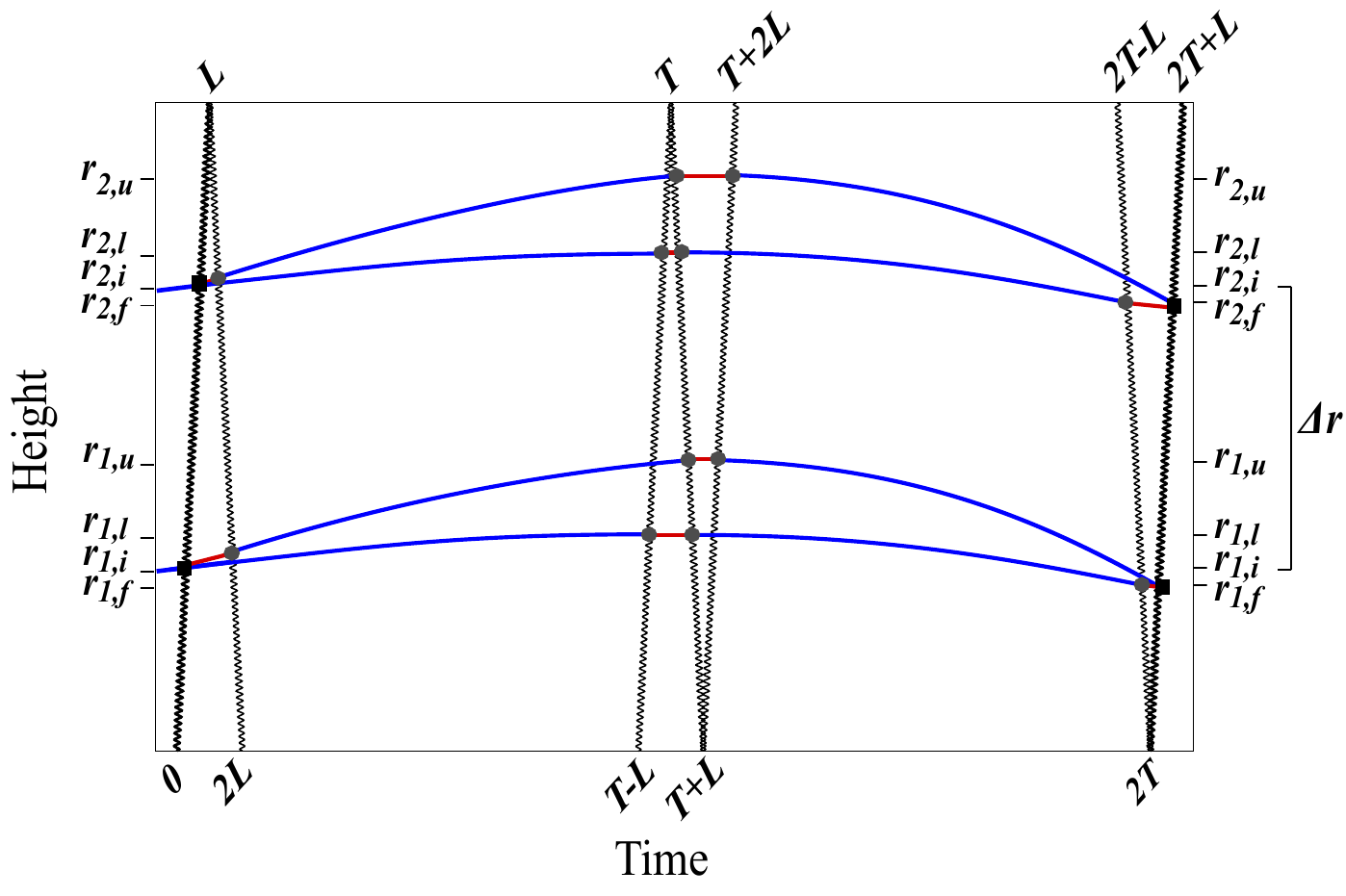}
        \caption{Schematic spacetime diagram of the atom gradiometer concept with $n=2$ large momentum transfer (LMT) kicks. The  experiment consists of a pair of cold-atom interferometers performing single-photon transitions between a ground state (blue) and an excited state (red). Height, or radial position, is shown on the vertical axis, and the time axis is horizontal. The laser pulses (wavy lines) traveling across the baseline from opposite sides are used to divide, redirect, and recombine the atomic states, yielding interference patterns that are sensitive to the modulation of the atomic transition frequency caused by couplings to DM. Beamsplitter ($\pi/2$) pulses, which begin and end the sequence, are shown in black, while $\pi$-pulses are shown in light grey. Atom-light interactions with $\pi/2$-pulses are shown as black squares, while atom-light interactions with $\pi$-pulses are shown with grey circles. The gradiometer length $\Delta r$ is also shown.}
    \label{fig:space_time}
\end{figure}

Using Eq.~\eqref{integral1} and summing over all paths in which the atom is in the excited state, we can find the total ULDM-induced phase difference for a single-photon atom interferometer labelled by $p$. Following a sequence as depicted in Fig.~\ref{fig:space_time} (which is for $n=2$), the phase of the $p^{\rm{th}}$ interferometer is
\begin{equation}
\begin{split}
\Phi_p = \sum_{m=1}^{n/2}  &\left[ \Phi_{T-(n-2m+1)L+r_{p,l}}^{T-(n-2m-1)L-r_{p,l}}+\Phi_{2T-(n-2m)L-r_{p,f}}^{2T-(n-2m)L+r_{p,f}} \right.\\ 
&\left. -\Phi_{T+(2m-1)L-r_{p,u}}^{T+(2m-1)L+r_{p,u}}-\Phi_{(2m-2)L+r_{p,i}}^{2mL-r_{p,i}} \right]\;,
\end{split}
\label{full signal AI}
\end{equation}
where $n$ is the number of large momentum transfer kicks (LMT)~\cite{Rudolph:2019vcv},
$L$ is the baseline separation between the laser and the mirror, and $T$ is the interrogation time, which is approximately half the duration of a single interferometric sequence that lasts for a total time $2T+L$.
With our convention, there are $4n-1$ pulses in total, of which two are
$\pi/2$-pulses and the remaining $4n-3$ are $\pi$-pulses.
These are distributed such that at the beginning and the end of the sequence, there are $(n-1)$ $\pi$-pulses, while around time $T$ in the middle of the sequence there are $(2n-1)$ $\pi$-pulses (the `mirror sequence') with $n-1$ $\pi$-pulses before time $T$ and $n$ $\pi$-pulses after.
With our labelling convention, the $\pi/2$-pulses are emitted at times $0$ and $2T$.
 The position vectors label the path taken by the excited state of the $p$\textsuperscript{th} interferometer: $r_{p,i}$, $r_{p,l}$, $r_{p,u}$ and $r_{p,f}$ refer to the initial, lower, upper and final path of the $p^{\mathrm{th}}$ interferometer as viewed in the laboratory frame, respectively. Finally, the sum is over positive integers, as the number of LMT kicks, $n$, is even to ensure that the atom is in the ground state along the longest paths.
 
The ULDM signal for the atom gradiometer corresponds to the difference between the total phase differences of two atom interferometers. Using Eq.~\eqref{full signal AI}, the atom gradiometer signal is defined as
\begin{equation}
\Phi_{\rm{s}} = \Phi_1-\Phi_2\;.
\label{full signal}
\end{equation}
Since $\omega_{\phi} L \ll 1 $ and in the limit that the launch velocity is the same in both interferometers, we show in Appendix~\ref{A1} that
Eq.~\eqref{full signal} can be reduced to the relatively compact expression: 
\begin{equation}
\begin{split}
\Phi_{\mathrm{s}} = \frac{\Delta r}{L} \Big\{ \Big[ & \Phi^{T+L}_{T-(n-1)L} - \Phi^{nL}_{0} \Big]  \\
 \qquad & -  \Big[ \Phi^{2T+L}_{2T-(n-1)L} - \Phi^{T+nL}_{T} \Big] \Big\}\;,
\end{split}
\label{approximate signal amplitude}
\end{equation}
 where we defined
 \begin{equation}\label{eq:deltar_defintion}\Delta r = r_{2,i}-r_{1,i}=r_{2,u}-r_{1,u}=r_{2,l}-r_{1,l}= r_{2,f}-r_{1,f}
 \end{equation}
 as the gradiometer length. Having the same launch velocity in both interferometers ensures that all of the equality signs hold in Eq.~\eqref{eq:deltar_defintion}. We can also obtain a compact expression for the signal amplitude (again, see Appendix~\ref{A1} for details): 
\begin{equation}
\begin{split}
\overline{\Phi}_{\rm{s}}=  8 \frac{\overline{\Delta \omega_{A}}}{m_{\phi}}\frac{\Delta r}{L} & \Bigg |\sin \left[\frac{m_{\phi}nL}{2}\right] \sin \left[\frac{m_{\phi} T}{2}\right] \\
&\quad \times \sin \left[\frac{m_{\phi}(T-(n-1)L)}{2}\right]  \Bigg |\;.
\label{signal amplitude}
\end{split}
\end{equation}
 In this expression, we have neglected the sub-leading kinetic corrections to the DM angular frequency, so that $\omega_{\phi}= m_{\phi}$.

 Setting $\Delta r = L$ in Eq.~\eqref{approximate signal amplitude} and Eq.~\eqref{signal amplitude} we recover the result presented in Ref.~\cite{Arvanitaki:2016fyj}, in which the authors considered experimental configurations in which two interferometers are separated by a distance that is comparable to the length of the baseline (i.e.\ $\Delta r \approx L$). While this approximation is valid for long-baseline setups, the $\Delta r/L$ factor presented in this work becomes an important correction to the ULDM signal in compact atom gradiometers. For example, in the long-baseline setup considered in Ref.~\cite{Arvanitaki:2016fyj}, $\Delta r/L=\SI{980}{m}/\SI{1000}{m}=0.98$, whereas for a compact atom gradiometer such as AION-10, $\Delta r/L = \SI{5}{m}/\SI{10}{m}=0.5$.
 
For a compact gradiometer where $nL \ll T$ and $m_{\phi} n L\ll1$, it is a good approximation to simplify Eq.~\eqref{signal amplitude} further to $\overline{\Phi}_{\rm{s}}\approx 4\,\overline{\Delta \omega_A}\, n \Delta r\, \sin^2\left[m_{\phi}T/2 \right]$. This approximate form has the advantage that it is relatively straightforward to determine the ULDM DM mass that maximises $\overline{\Phi}_{\rm{s}}$. Although it naively seems that this should occur at $m_\phi= \pi/T$, this is not correct as there is an additional mass term implicit in $\overline{\Delta \omega_A}$ (recall from Eq.~\eqref{signal amplitude 2} that $\overline{\Delta \omega_A} \propto m_{\phi}^{-1})$. When taking this additional mass-term into account, we find that $\overline{\Phi}_{\rm{s}}$ is maximised for $m_{\phi} \approx 2.33/T$.

\section{ Sensitivity Dependence on Tuneable Experimental Parameters \label{sec:sensitivityscaling}}

In the previous section, we obtained in Eq.~\eqref{signal amplitude} a compact expression for the signal amplitude
from a single interferometric sequence that lasts for a time $2T+L$. In a realistic experiment, this sequence will be repeated
many times giving rise to $N$ measurements over the course of a measurement campaign set by the integration time $T_\mathrm{int}$.
Furthermore, the measured signal will consist of a contribution from background noise sources in addition to any ULDM-induced phase.
In this section, we provide a general discussion of how the sensitivity to ULDM parameters scales with tuneable experimental parameters
in a compact atom gradiometer. We then apply the results to the AION-10 setup.

To extract information from a discrete series of measurements collected at different points in time, the estimator of the power spectral density (PSD) is commonly used (see e.g.~\cite{VanderPlas_2018}). The PSD is useful because it contains information on the frequency spread and amplitude of the analysed data. 
We define $\Phi_m$ as the phase measured from a gradiometer sequence that begins at time $m \Delta t$, where $\Delta t$ is
the temporal separation between successive measurements (i.e.\ $1/\Delta t$ is the sampling rate),
and we assume that measurements are made continuously for the total integration time, i.e., $T_\mathrm{int} = N \Delta t$.
By first taking the discrete Fourier transform $\tilde{\Phi}_k$ of the measured signal, defined as
\begin{equation}
\tilde{\Phi}_k = \sum_{m=0}^{N-1} \Phi_m \exp{\left(-\frac{2 \pi i m k}{N} \right)}\;,
\end{equation}
where $k$ is an integer taking values from $0$ to $N-1$,
we can define the PSD~as
\begin{equation}\label{eq:PSD}
S_k = \frac{(\Delta t)^2}{T_{\rm{int}}}  \left| \tilde{\Phi}_k  \right|^2\;.
\end{equation}

We will use the signal-to-noise ratio (SNR) 
as an estimator of the signal strength. We define the angular frequency $\omega = 2 \pi k / T_{\rm{int}}$, such that 
the SNR at the frequency $\omega_0$ takes the form
\begin{equation}
    \mathrm{SNR} = \frac{S_\mathrm{s} (\omega_0)}{\sigma_{S_\mathrm{n}}(\omega_0)}\;,
    \label{SNR}
\end{equation} 
where $S_\mathrm{s}(\omega)$ is the PSD of the ULDM signal and $\sigma_{S_\mathrm{n}}(\omega)$ is the standard deviation of the noise PSD.

Although several background components will contribute to the noise PSD $S_\mathrm{n}(\omega)$ (see for example the detector systematics discussion in Ref.~\cite{Coleman:2018ozp}),
the design goal of an experiment's detection system is that the dominant phase noise is from atom shot noise. This is a challenging goal as reaching the shot noise limit has so far proved elusive except in smaller atom interferometers.
For a single interferometer, the noise variance is $\sigma^2= (C^2 N_a)^{-1}$ for phase differences close to $\pi/2$, where $N_a$~is the number of atoms in the cloud and $C \leq 1$ is the interferometer contrast~\cite{Le_Gou_t_2008, PhysRevA.47.3554}. 
The interferometer contrast~$C$ is important because it characterises the amplitude of the oscillation of the number of atoms in the ground/excited state, from which the interferometer phase is inferred (see e.g.~\cite{Roura:2015xsa}).
By the Wiener-Khinchin theorem, this implies that the noise PSD for a single interferometer is $\Delta t / C^2 N_a$, while for an atom gradiometer employing identical interferometers, $S_\mathrm{n} = 2\Delta t/ C^2 N_a$. Because atom shot noise is white noise, the standard deviation of the gradiometer noise PSD is 
 \begin{equation}
     \sigma_{S_\mathrm{n}} = S_\mathrm{n} = \frac{2\Delta t}{ C^2 N_a}\;,
 \end{equation}
which is frequency-independent by definition. Hence, at $\mathrm{SNR}=1$ the peak of the ULDM-induced PSD is within one standard deviation of the mean of the noise PSD.

Next, we consider the PSD for the ULDM signal, $S_\mathrm{s} (\omega)$. From the PSD defining equation, Eq.~\eqref{eq:PSD}, we can see that the maximum of $S_\mathrm{s} (\omega)$ will be proportional to $\overline{\Phi}_{\rm{s}}^2$. The frequency spread of $S_\mathrm{s} (\omega)$ is related to the reciprocal of the coherence time,  $2\pi/\tau_c  \sim 10^{-7}\, (m_{\phi}/10^{-15}~\mathrm{eV}) ~\mathrm{Hz}$, while the experiment's frequency resolution is given by $2\pi/T_\mathrm{int}$.  

In the regime $T_\mathrm{int} < \tau_c$, the ULDM signal is dominated by a single frequency, which implies that the PSD is approximately  $S_\mathrm{s}(\omega_0) \approx T_\mathrm{int} |\overline{\Phi}_{\rm{s}}|^2$ at a frequency $\omega_0$ largely set by $m_{\phi}$, and zero elsewhere.   
Hence, the experiment's sensitivity to the coupling constants increases with integration time and is given by
\begin{equation}
d_\phi \simeq \frac{\sqrt{\mathrm{SNR}}}{\overline{\Phi}_{\rm{R}}} \times \sqrt{\frac{S_n}{T_\mathrm{int}}}\;,
\end{equation}
where in order to isolate the dependence on the couplings $d_{m_e}$ and $d_e$, we have defined
\begin{align}
\overline{\Phi}_{\rm{s}} &= d_{\phi} \overline{\Phi}_{\rm{R}}\\
d_\phi&= d_{m_e} + (2+\xi_A) d_e\;,
\end{align}
where $\overline{\Phi}_{\rm{R}}$ is the remaining part of the signal amplitude with the couplings factored out.

In the limit $T_\mathrm{int} > \tau_c$, it might naively be expected that the experiment's sensitivity to $d_\phi$ follows a similar argument. 
However, this is not the case since the PSD of the entire signal will no longer correspond to a spike in frequency space, but will have a finite width and profile dictated by the DM velocity distribution. In this case, a likelihood profile analysis could be used to extract the experiment's sensitivity to~$d_\phi$ (see e.g.~\cite{Foster:2017hbq}). Alternatively, Bartlett's method~\cite{VanderPlas_2018, Budker:2013hfa} can be applied to find individual PSDs from data streams of duration $\tau_{\rm{Bart}}\lesssim \tau_c$, which are then averaged.
 An advantage of Bartlett's method over a likelihood-based analysis is that the detailed form of the DM speed distribution does not need to be specified. 
Bartlett's method reduces the frequency resolution of the signal PSD to a spike, while also reducing the standard deviation of the noise PSD by a factor $\sqrt{T_\mathrm{int}/\tau_{\rm{Bart}}}$. 
In the limit $\tau_{\rm{Bart}} = \tau_c$, the experiment's sensitivity to the couplings strengths is therefore given~by
\begin{equation}
d_\phi \simeq \frac{\sqrt{\mathrm{SNR}}}{\overline{\Phi}_{\rm{R}}} \times \sqrt{\frac{S_n}{\sqrt{\tau_c T_\mathrm{int}}}}\;.
\label{scaling}
\end{equation}

For terrestrial atom gradiometers, Newtonian gravity gradient noise (GGN) is expected to exceed atom shot noise at frequencies less than approximately $\SI{e-1}{Hz}$, which corresponds to a mass $m_{\phi}\approx 4\times 10^{-16}$~eV.  If the GGN noise cannot be mitigated, this will impose a lower limit on the frequencies that can be probed~\cite{Arvanitaki:2016fyj}. In our treatment, we take a conservative approach and only show projections above $\SI{e-1}{Hz}$.
 For compact atom gradiometers that operate on the time scale of years (i.e.\ $T_\mathrm{int}\gtrsim \mathrm{few}\times 10^7~\mathrm{s}$), the total integration time will in general exceed the coherence time of an ULDM signal for all ULDM masses of interest.
Hence, the scaling of the experimental parameters for a linear scalar ULDM-electron or photon interaction is described by Eq.~\eqref{scaling}. 

Pulling all parts of this discussion together, we find that the experiment's maximum sensitivity to a linear scalar ULDM-electron or photon interaction (in the limit  $T_\mathrm{int} > \tau_c$) scales with experimental parameters in the following way:
\begin{equation}
    d^{\mathrm{best}}_\phi \sim \Bigg (\frac{1}{T}\Bigg)^{5/4}\frac{1}{C\, n \Delta r} \, \,\Bigg ( {\frac{\Delta t}{N_a}} \Bigg ) ^{1/2} \Bigg (\frac{1}{T_\mathrm{int}}\Bigg ) ^{1/4} \;.
\label{scaling2}
\end{equation}
The additional $m_{\phi}$ dependence in $\tau_c$ has two effects; the maximum sensitivity occurs for ULDM masses $m_{\phi}\approx 2.04/T$, which is slightly below $m_{\phi}\approx2.33/T$ (where $\overline{\Phi}_s$ is maximised), and the scaling is $T^{-5/4}$ rather than $T^{-1}$.
Equation~\eqref{scaling2} reveals a hierarchy of importance amongst the tuneable experimental parameters: the sensitivity scales as $T^{-5/4}$, so is most sensitive to changes in this parameter; it varies linearly with the inverse of $C$, $n$, and $\Delta r$; with the square root of $\Delta t$ and $N_a$; and with the quartic root of $T_{\rm{int}}$, indicating the least sensitivity to this parameter. 

\begin{figure*}[t]
\subfigure{\includegraphics[width=0.47\textwidth]{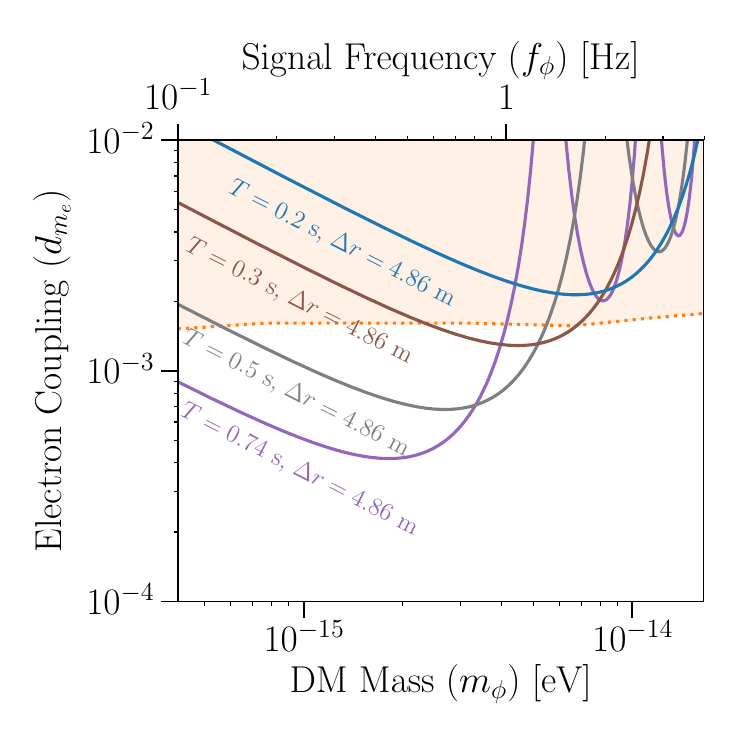}}
    \subfigure{\includegraphics[width=0.47\textwidth]{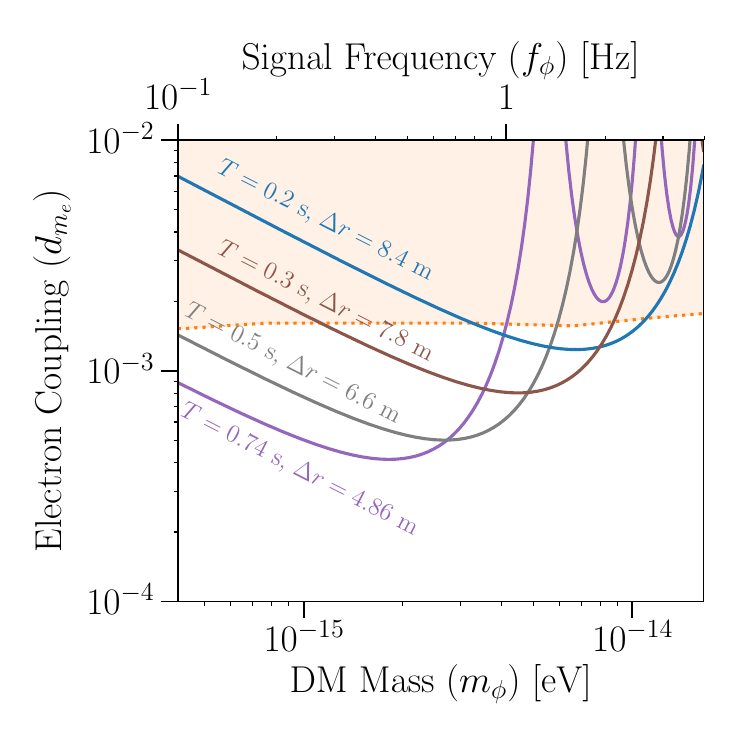}}
    \caption{Estimated reach in the shot noise limit of searches for scalar DM interactions with electrons within the $\SI{0.1}{Hz} \leq f_\phi \leq \SI{4}{Hz}$ signal window. We plot curves assuming $\mathrm{SNR}=1$ and an integration time of $T_\mathrm{int}=10^8~\!\rm{s}$ for a \SI{10}{m} atom gradiometer with $n=1000$, $S_n = 10^{-8}~\!\mathrm{Hz}^{-1}$, and: fixed gradiometer length $\Delta r=\SI{4.86}{m}$ for different interrogation times $T$ (left panel); optimal gradiometer length $\Delta r$ for different interrogation times $T$ (right panel). As $T$ increases, the curves move to higher values of $m_{\phi}$ and $d_{m_e}$. The loss in sensitivity can be mitigated somewhat by increasing $\Delta r$.
    The orange regions show parameter space that has already been excluded through searches for violations of the equivalence principle via terrestrial torsion balance experiments~\cite{Wagner:2012ui} and MICROSCOPE~\cite{Berge:2017ovy}.}
    \label{fig:sensitivity max minimum}
\end{figure*}

The scaling in Eq.~\eqref{scaling2} shows that the interrogation time $T$ is important as it not only sets the ULDM mass at which the experiment has the maximum sensitivity, but
it also affects the experiment's maximum sensitivity reach in $d_{\phi}$.
We demonstrate this explicitly in the left panel of Fig.~\ref{fig:sensitivity max minimum}, where we have plotted the electron coupling when $\mathrm{SNR}=1$ for different values of the 
the interrogation time from \SI{0.74}{s} to \SI{0.2}{s} while keeping the gradiometer length ($\Delta r$) and all other parameters fixed. 
The left panel shows that as $T$ increases, the curves both move to higher ULDM masses  \textit{and} higher values of $d_{m_e}$ as predicted by Eq.~\eqref{scaling2}.
Hence, shorter interrogation times imply a loss in sensitivity. In these plots, we have assumed $T_{\rm{int}}=10^8~\!\mathrm{s}$ to employ Bartlett's method, and set $S_n = 10^{-8}~\!\mathrm{Hz}^{-1}$ and $n=1000$. As we will discuss in more detail in the following section, these values are indicative of the parameters ultimately envisaged for an atom gradiometer such as AION-10.  
 
 In the right panel, we show that this loss in sensitivity can be mitigated somewhat by increasing the gradiometer length. However, there is a practical limit on how large the gradiometer length can be, which is provided by the baseline of the experiment. In the right panel, we have chosen $\Delta r$ to be as large as possible while ensuring that the atoms remain inside a $\SI{10}{m}$ atom gradiometer at all times.
 
 Finally, we see that there may be some benefit in running an experiment with different values of~$T$. The oscillatory nature of the signal amplitude $\overline{\Phi}_{\rm{s}}$ means that for fixed~$T$, there will be points in parameter space where the amplitude is significantly reduced (e.g., at $m_{\phi}\approx 6\times 10^{-15}~\!\mathrm{eV}$ when $T=\SI{0.74}{s}$). By running with a different $T$ value, for instance, $T=\SI{0.3}{s}$, the gaps in sensitivity can be closed. To maximise the experiment's reach, it may even be desirable to employ one baseline to run two atom interferometers with alternate launch conditions, allowing different regions of the signal frequency to be probed simultaneously. 

\section{Optimising for AION-10 \label{sec:optAION}}

\begin{figure*}[t!]
    \subfigure{\includegraphics[width=0.47\textwidth]{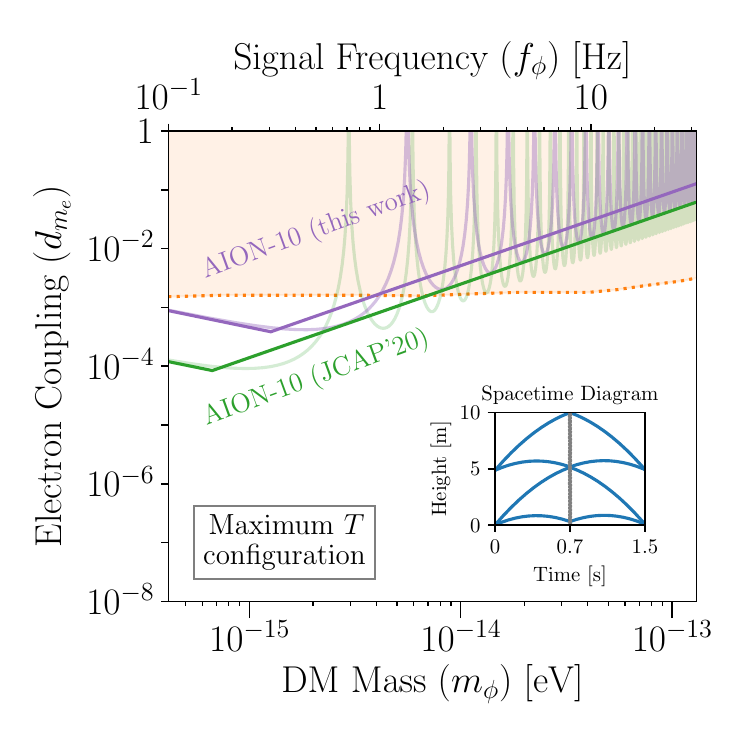}}
    \subfigure{\includegraphics[width=0.47\textwidth]{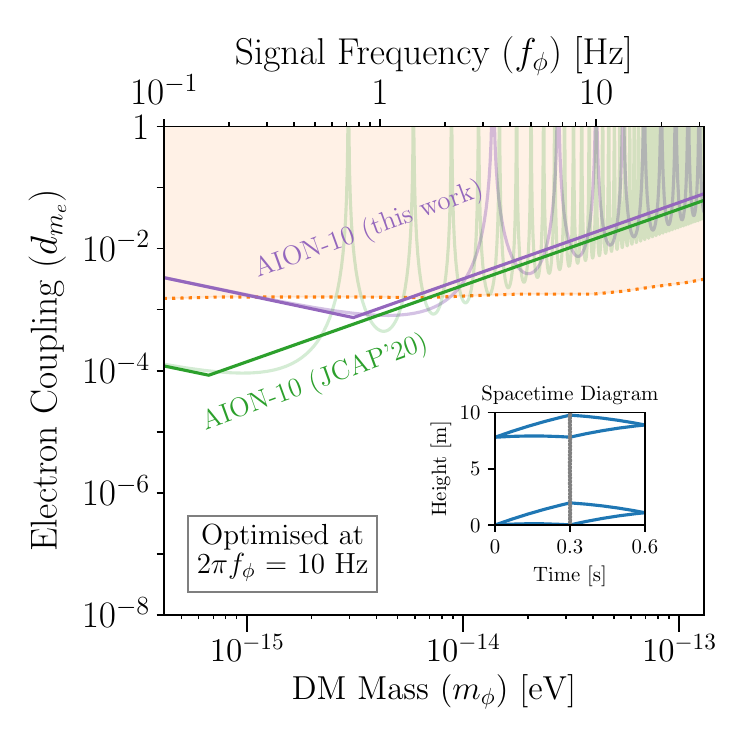}}
    \caption{ Estimated reach in the shot noise limit of searches for scalar DM interactions with electrons using a \SI{10}{m} atom gradiometer. We assume $T_\mathrm{int}=10^8~\!\mathrm{s}$,  $S_{n}=10^{-8}~\!\mathrm{Hz}^{-1}$ and $\mathrm{SNR}=1$.
    In purple, `AION-10 (this work)', we plot the result from the refined AION-10 calculation. The bold lines show the power-averaged envelope, while the fainter curves show the full estimated projection.
   Left panel:  the experimental parameters are $n=1000$, $\Delta r=\SI{4.86}{m}$, $r_{1,i}=\SI{0}{m}$ and $v_i= \SI{4.04}{m/s}$ resulting in $T = \SI{0.74}{s}$, which is the maximum value of $T$ in a \SI{10}{m} gradiometer.
    Right panel: the parameters are $n=1000$, $\Delta r=\SI{7.8}{m}$, $r_{1,i}=\SI{0}{m}$, $v_i= \SI{1.5}{m/s}$ and $T = \SI{0.3}{s}$, which are chosen to maximise the sensitivity at $2\pi f_\phi = \SI{10}{Hz}$.
    The orange regions show parameter space that has already been excluded through searches for violations of the equivalence principle via terrestrial torsion balance experiments~\cite{Wagner:2012ui} and MICROSCOPE~\cite{Berge:2017ovy}. 
    In green, `AION-10 (JCAP'20)', we plot the `goal' estimate from Ref.~\cite{Badurina:2019hst}.  The insets in each plot depict the spacetime diagram of the interferometric sequence for the aforementioned set of experimental parameters.}
    \label{fig:sensitivity}
\end{figure*}

To maximise a potential ULDM-induced signal in a single-photon atom gradiometer with frequency $\gtrsim \SI{e-1}{Hz}$, a careful selection of experimental parameters is required. This may at first seem trivial given that Eq.~\eqref{scaling2} suggests that the experiment's sensitivity is maximised at $m_\phi \approx 2.04/T$ for high sampling rates, large atom cloud populations, high LMT, high contrast, large separations between atom interferometers, and long interrogation and integration times. However, the experimental sensitivity also indirectly depends on the initial launch velocity ($v_i$) and the position from which the atom clouds are initially launched ($r_{1,i}$ and $r_{2,i}$, following the notation used in Fig.~\ref{fig:space_time}).
In our analysis, we require that the paths of different interferometers do not overlap as this completely avoids the possibility of collisional losses. This requirement, together with the constraint that atomic states must be confined within the baseline of the experimental apparatus, implies that the interrogation time $T$ becomes a function of the other experimental parameters governing the interferometric sequence.
Although the AION-10 design is not yet final, we will use experimental parameters that are indicative of the ambitious goals envisaged in the final phase of AION-10
to describe how the search for optimal AION-10 experimental parameters can be carried out.

For a baseline $L=\SI{10}{m}$ and assuming that $n=1000$, we carried out the search for optimal parameters by scanning over values of $v_{i}$, $r_{1,i}$ and $r_{2,i}$. We assume that both clouds are launched with the same velocity and that the atoms are subject to a constant gravitational acceleration $g\simeq 9.81~\!\mathrm{m}/\mathrm{s}^2$. In calculating atom trajectories, we have also assumed that all of the momentum transfer from the $n$~LMT kicks is transmitted instantaneously while in reality this would occur over a timescale $\mathcal{O}(500~\!\mathrm{\mu s})$~\cite{Rudolph:2019vcv}. However, this gives a negligible correction to the trajectories over timescales $\sim T$. 

There is a configuration that maximises $T$ where analytic results can be obtained. In this configuration,
the lower interferometer is positioned at $r_{i,1}=0$ and the atoms start and end at $r_{1,f}=0$, which implies that the sequence is symmetric around~$T$.
The upper interferometer will start at $r_{2,i}=\Delta r$, and its upper arm is made to reach $r_{2,u}=L$ after a time~$T$.
Finally, the upper arm of the lower interferometer is made to reach the same position as the lower arm of the top interferometer at time~$T$, i.e.,~$r_{1,u}=r_{2,l}$.
This configuration occurs if $2 v_i = \sqrt{8 g L + 9 v_{n}^2} - 4 v_n$, $g T=v_i+ v_n/2$ and $\Delta r = v_n T$, where $v_n = n \hbar k_{A}/m_{A}$ is the speed imparted to the $^{87}\mathrm{Sr}$ atom after $n$~LMT kicks, which depends on the angular wavenumber $ k_{A}$ of the clock transition and the $^{87}\mathrm{Sr}$ atom mass $m_A$. 
The estimated projection in the shot noise limit when $\mathrm{SNR}=1$ for the maximum $T$ configuration is shown by the purple line in the left panel of Fig.~\ref{fig:sensitivity}.
For $L=\SI{10}{m}$ and $n=1000$, we find that $T = \SI{0.74}{s}$, $v_i= \SI{4.04}{m/s}$ and $\Delta r=\SI{4.86}{m}$.
The inset shows the spacetime diagram of the interferometric sequence and confirms that the trajectories remain within the baseline at all times during the sequence.
We have also checked numerically that this configuration results in the maximum signal amplitude $\Phi_{\rm{s}}$.

In calculating the purple line in the left panel of Fig.~\ref{fig:sensitivity}, we have assumed $T_{\rm{int}}=10^8~\!\mathrm{s}$ to employ Bartlett's method
with a gradiometer's noise power spectral density $S_{n}=10^{-8}~\!\mathrm{Hz}^{-1}$.
The plots and troughs of the curves that were apparent in Fig.~\ref{fig:sensitivity max minimum} are again found in Fig.~\ref{fig:sensitivity} and are shown by the lighter purple lines.
To facilitate the comparison with other results in the literature, the darker, straighter lines show the power-averaged envelope using the approximation $|\sin x|=\min\{x,1/\sqrt{2}\}$ in Eq.~\eqref{signal amplitude}, which smooths out the peaks and troughs arising from the fast oscillation of the trigonometric functions. However, it should be understood that this approximation is a visual aid and that limits display the peak and trough structure.

The orange shaded regions in Fig.~\ref{fig:sensitivity} show parameter space that has already been excluded through searches from terrestrial torsion balance experiments~\cite{Wagner:2012ui} and MICROSCOPE~\cite{Berge:2017ovy} so we see that with these parameter values, there is the 
possibility of probing unconstrained regions of parameter space.
We plot results for the DM mass range $\SI{e-15}{eV} \lesssim m_\phi \lesssim \SI{e-13}{eV}$ but they could in principle be extended to even larger masses. While the lower bound is set by the frequency at which gravity gradient noise dominates, the upper bound corresponds to the highest frequency at which the DM wave remains coherent over successive measurements, such that $\tau_c\gtrsim \Delta t$~\cite{Arvanitaki:2016fyj,Derevianko:2016vpm}. For a fiducial value $\Delta t \simeq \SI{10}{s}$, this corresponds to $m_\phi \lesssim \SI{e-9}{eV}$.

The green lines in Fig.~\ref{fig:sensitivity} show the estimates based on a previous set of calculations in Ref.~\cite{Badurina:2019hst}. Unlike in this paper, the calculations in Ref.~\cite{Badurina:2019hst} did not include the $\Delta r/L$ correction in Eq.~\eqref{signal amplitude} and used a larger value of $T$ that was not consistent with the requirement that the atom trajectories remain within the baseline at all times. The result is that the refined calculations shown by the purple line  is at larger values of the electron coupling.

While the results in the left panel of Fig.~\ref{fig:sensitivity max minimum} show the estimated projection for the maximum $T$ configuration, it is possible to arrange for different sequences.
In particular, a different launch scheme could be employed to focus on different regions of parameter space.
In the right panel of Fig.~\ref{fig:sensitivity max minimum}, for example, we show a configuration that maximises the sensitivity at a signal angular frequency of \SI{10}{Hz}. This means that we have found the experimental parameters that result in the lowest electron coupling at this frequency (corresponding to a mass $m_{\phi}\simeq 7\times10^{-15}~\!\mathrm{eV}$). This configuration was found by scanning over values of $v_{i}$, $r_{i,1}$ and $r_{i,2}$, resulting in $T = \SI{0.3}{s}$, $\Delta r=\SI{7.8}{m}$, $r_{1,i}=\SI{0}{m}$ and $v_i= \SI{1.5}{m/s}$. The spacetime diagram of the interferometric sequence is shown in the right-panel inset, and shows that in this case, the trajectories remain inside the baseline at all times but the sequence ends at a different position compared to the launch position. Comparing the left and right panels of Fig.~\ref{fig:sensitivity max minimum} we see that the power-averaged envelope has shifted to larger frequencies so that the refined calculation in the right panel is closer to the older result from Ref.~\cite{Badurina:2019hst} over a larger region of parameter space. However, this has come at the cost of reduced sensitivity to signal frequencies in the range from 0.1~Hz to 1~Hz.

\begin{figure}[t]
    \centering    
    \includegraphics[width=0.49\textwidth]{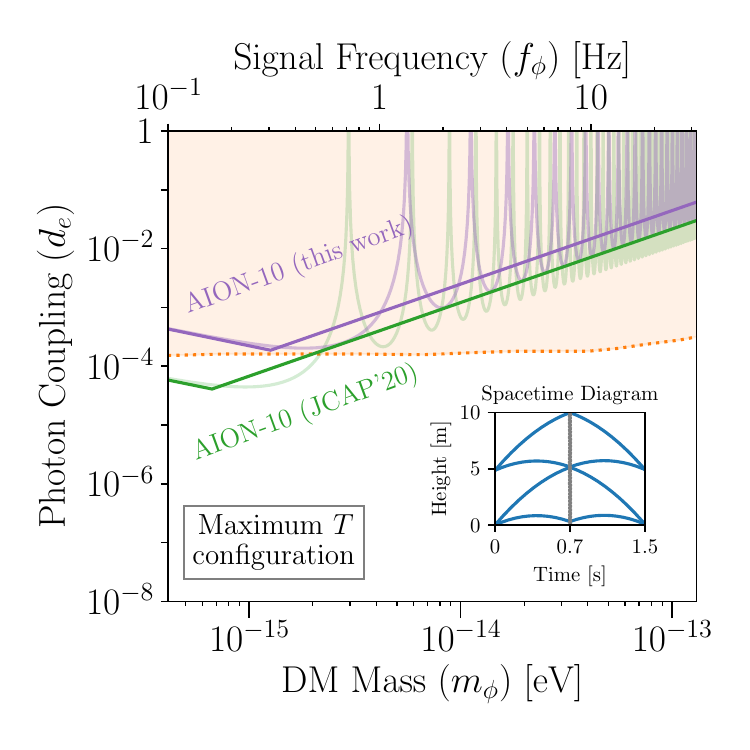}
    \caption{Estimated reach in the shot noise limit of searches for scalar DM interactions with photons using a \SI{10}{m} atom gradiometer after an integration time of $T_\mathrm{int}=10^8~\!\mathrm{s}$, assuming $S_{n}=10^{-8}~\!\mathrm{Hz}^{-1}$ and $\mathrm{SNR}=1$. In purple, `AION-10 (this work)', we plot the refined projection for AION-10 for
    the maximum $T$ parameters $n=1000$, $\Delta r=\SI{4.86}{m}$, $r_{1,i}=\SI{0}{m}$ and $v_i= \SI{4.04}{m/s}$ resulting in $T = \SI{0.74}{s}$. In green, `AION-10 (JCAP'20)', we plot the AION-10 `goal' estimate from Ref.~\cite{Badurina:2019hst}.   The bold lines show the power-averaged sensitivity, while the fainter curves show the full expected sensitivity.
     The orange regions show parameter space that has already been excluded through searches for violations of the equivalence principle via terrestrial torsion balance experiments~\cite{Wagner:2012ui} and MICROSCOPE~\cite{Berge:2017ovy}.  The inset depicts the spacetime diagram of the interferometric sequence.}
    \label{fig:photon sensitivity}
\end{figure}

As shown in the insets in Fig.~\ref{fig:sensitivity}, the atom trajectories display differences depending on the angular frequency at which the sensitivity is optimised. 
At larger $T$ (or equivalently, when optimising at lower frequencies), the area enclosed by the atom paths is maximised. 
This will favour schemes in which the interferometers are separated by $\sim L/2$ and the atoms in each travel a maximum displacement of $\sim L/2$. 
On the contrary, at smaller $T$ (optimising at higher frequency), the separation between the interferometers is maximised. This will favour schemes in which the interferometers are separated by $\sim L$. 
In this case, operating a further atom interferometer at $\sim L/2$ could be envisaged, as this would provide a further independent measurement of the DM field and may lead to better control of the GGN.

For completeness, in Fig.~\ref{fig:photon sensitivity} we also present the projected gradiometer's optimal sensitivity to scalar DM coupled to photons for the maximum $T$ configuration, which optimises the sensitivity at lower values of $m_{\phi}$. The experimental parameters are the same as those used in the left panel of Fig.~\ref{fig:sensitivity}. As in Fig.~\ref{fig:sensitivity}, we again find that the refined calculation in this work is at higher values of the coupling compared to the calculation in Ref.~\cite{Badurina:2019hst}. For the photon coupling, the constraints from atom interferometery are less stringent than the space-based constraints from MICROSCOPE but have the potential to match the sensitivity from terrestrial torsion balance experiments. 

\subsection{Resonant mode with multiple spacetime diamonds \label{sec:resonant}}

\begin{figure}[t!]
    \centering    
    \includegraphics[width=0.47\textwidth]{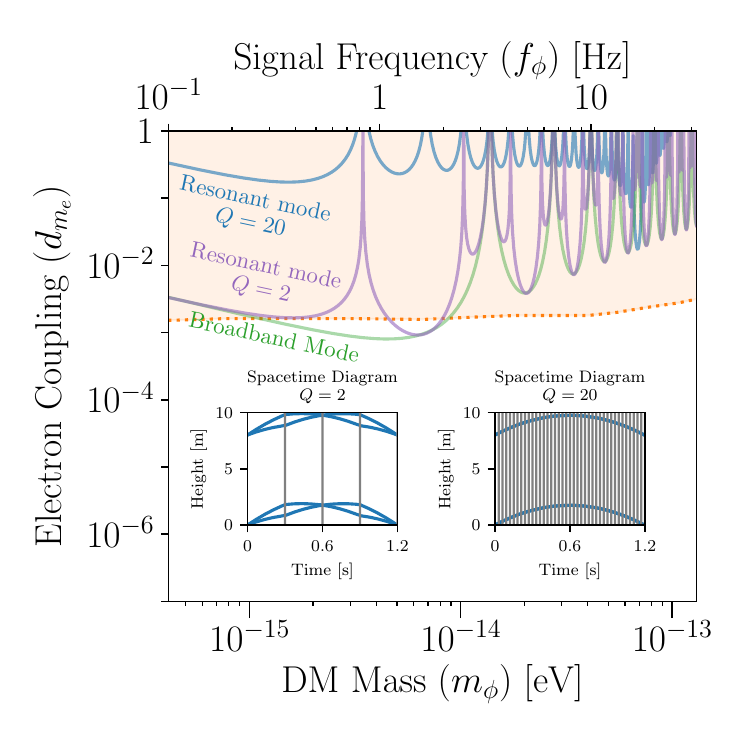}
    \caption{Comparison of the estimated reach in the shot noise limit of searches for scalar DM interactions with electrons between broadband and resonant modes after $T_\mathrm{int}=\SI{e8}{s}$ and assuming $S_{n}=10^{-8}~\!\mathrm{Hz}^{-1}$ and $\mathrm{SNR}=1$.
    The green curve shows the broadband mode estimate, previously shown as the `AION-10 (this work)' curve in the right panel of Fig.~\ref{fig:sensitivity}.
    The purple curves shows the resonant mode projection for
    $Q = 2$, $T = \SI{0.3}{s}$, $n=500$, $\Delta r=\SI{8.0}{m}$, $r_{1,i}=\SI{0.0}{m}$ and $v_i= \SI{4.25}{m/s}$. The blue curve shows the resonant mode projection for $Q = 20$, $T = \SI{0.03}{s}$, $n=50$, $\Delta r=\SI{8.0}{m}$, $r_{1,i}=\SI{0.0}{m}$ and $v_i= \SI{5.7}{m/s}$.
    The insets depict the spacetime diagrams for each resonant mode sequence. The orange regions show the excluded parameter space from torsion balance experiments~\cite{Wagner:2012ui} and MICROSCOPE~\cite{Berge:2017ovy}.
    }
    \label{fig:resonant sensitivity}
\end{figure}

All of the results so far have used the sequence shown in Fig.~\ref{fig:space_time}, which consists of a single closed spacetime 'diamond'. We will refer to this sequence as the `broadband mode' owing to the sensitivity over a wide range of frequencies. However, an alternative sequence that enhances the sensitivity at certain frequencies could also be employed~\cite{Graham:2016plp}. This `resonant mode' scheme consists of $Q$ closed spacetime diamonds each lasting for a time $\sim 2T$. Two examples of the closed spacetime diagrams for $Q=2$ and $Q=20$ are shown in the insets in Fig.~\ref{fig:resonant sensitivity}.
The resonant mode employs $n_{\rm{tot}}=2Q(2n-1)+1$ laser pulses in total, which consists of two $\pi/2$-pulses that define the start and end of the interferometric sequence and are emitted at times $0$ and $2QT$, and $\left[2Q(2n-1)-1\right]$~$\pi$-pulses. The number of LMT-kicks during a single diamond in the sequence is again denoted by $n$.
The signal amplitude induced by ULDM for resonant mode searches takes the form 
\begin{equation}
\begin{split}
    \overline{\Phi}^{\,Q}_{\rm{s}}  &= 8  \frac{ \overline{\Delta \omega_A}}{m_{\phi}} \frac{\Delta r}{L}  \Bigg | \sin\left[\frac{m_{\phi}n L}{2}\right] \sin\left[\frac{m_{\phi}T}{2}\right] \\  & \qquad \times \sin \left[\frac{m_{\phi}(T-(n-1)L)}{2} \right]  \frac{\sin\left[Q m_{\phi} T\right]}{\sin\left[m_{\phi} T\right]} \Bigg |.
\end{split}
\label{eq:resonant mode amp}
\end{equation}
The derivation of this expression is given in Appendix~\ref{A_diamonds}.
The proposed advantage of the resonant mode is that
there is an enhancement by a factor~$Q$ at $m_{\phi}= \pi/T$ over a mass range $\sim \pi/(QT)$ relative to the broadband search using the same set of experimental parameters.
 The additional $Q$ enhancement and the resonant behaviour arises from the $\sin\left[Q m_{\phi} T\right]/\sin\left[ m_{\phi} T\right]$ term in Eq.~\eqref{eq:resonant mode amp}, which was not present in Eq.~\eqref{signal amplitude}.
 
In Fig.~\ref{fig:resonant sensitivity} we compare the estimated projections from broadband and resonant mode searches for two different values of~$Q$. In this comparison, we assume that the maximum total number of laser pulses is $n_\mathrm{max} =  4000$ and, for each value of $Q$, we choose the largest even value of $n$ that satisfies the inequality $2Q(2n-1)+1 \leq n_\mathrm{max}$. 
This implies that as $Q$ increases, $n$ must decrease but the maximum signal amplitude
will remain similar since $\overline{\Phi}^{\,Q}_{\rm{s}}\propto n Q$.
The purple line in Fig.~\ref{fig:resonant sensitivity} shows the sensitivity curve for a resonant mode configuration operating with $Q = 2$, $T = \SI{0.3}{s}$, $n=500$, $\Delta r=\SI{8.0}{m}$, $r_{1,i}=\SI{0.0}{m}$ and $v_i= \SI{4.25}{m/s}$; in blue, we show the sensitivity curve for a resonant mode configuration operating with $Q = 20$, $T = \SI{0.03}{s}$, $n=50$, $\Delta r=\SI{8.0}{m}$, $r_{1,i}=\SI{0.0}{m}$ and $v_i= \SI{5.7}{m/s}$. The corresponding spacetime diagrams are show in the inset.
As in our previous calculations, we have assumed an integration time of \SI{e8}{s} and a noise power spectral density of $\SI{e-8}{Hz^{-1}}$.

In green, we show the broadband mode sensitivity that was shown previously (as the purple line) in the right panel of Fig.~\ref{fig:sensitivity}.
Comparing the broadband operation to resonant mode searches, we find that neither resonant mode configuration  substantially increases the sensitivity relative to the broadband mode.
 For $Q = 2$, the maximum sensitivity reach is approximately equal to the one available to broadband searches at integer multiples of $m_{\phi} \sim \pi/T$, but is smaller for other frequencies. One possible advantage of the $Q=2$ sequence is that it begins and ends at the same location, which was not the case for the broadband sequence (cf.~inset in Fig.~\ref{fig:sensitivity} right). This would be an advantage for an experimental configuration where the detection and readout optics are located at the same location (e.g.~at the sidearm base).
Higher $Q$ values result in a smaller value of $T$, which is why the best sensitivity for $Q=20$ is shifted to higher values of $m_{\phi}$. 
The $Q = 20$ curve also more clearly displays the resonant structure as the sensitivity away from $m_{\phi}= \pi/T$ is substantially weaker.
While the high-$Q$ mode allows for the possibility of probing a specific frequency, for compact gradiometers, the curve's minimum will in general lie above the exclusion limits from torsion balance experiments and MICROSCOPE. Given current exclusion limits,  we therefore find that no significant improvement can be found when operating a compact atom gradiometer in resonant mode instead of broadband mode.

\section{Conclusions \label{sec:conclusions}}

Single-photon atom gradiometry is a powerful experimental technique that can be used to search for oscillations of fundamental constants caused by ultra-light dark matter (ULDM). 
While the existing formalism for calculating the gradiometer sensitivity to ULDM is applicable to $\mathcal{O}(\mathrm{km})$-baseline atom gradiometers in which two interferometers are separated by a distance that is comparable to the length of the baseline,
the calculations presented in this work provide a careful treatment that is also applicable to gradiometers that have a more compact design. In particular, we have been careful to include the contribution to the ULDM-induced phase from all segments where the atom is in the excited state, thus finding an additional $\Delta r/L$ correction-term in Eq.~\eqref{signal amplitude}.
In addition, we have chosen physically realistic experimental parameters that ensure the atoms remain confined with the compact baseline at all times.

 Using these results, we have refined the sensitivity estimates for AION-10, a compact \SI{10}{m} gradiometer that will be operated in Oxford. Using the ambitious experimental parameters envisaged for the final phase of AION-10, and assuming that the dominant phase noise is from atom shot noise, we find that AION-10
 has the potential to probe currently unconstrained values of the electron-ULDM coupling for ULDM masses around $10^{-15}~\!\mathrm{eV}$ (cf.\ Figs.~\ref{fig:sensitivity max minimum} and~\ref{fig:sensitivity}).
 We have also provided a discussion of the how the sensitivity scales with tuneable experimental parameters. This allows for a relatively straightforward remapping of our projected sensitivities to other parameter choices so that if, for instance, AION-10 ultimately achieves $n=100$ LMT kicks instead of $n=1000$, or if the number of strontium atoms per cloud is $N_a=10^6$ instead of $N_a=10^8$, Eq.~\eqref{scaling2} implies that the estimated reach of searches for scalar DM coupled to electrons will be weaker by a factor of 10 relative to the projections shown in Figs.~\ref{fig:sensitivity max minimum} or~\ref{fig:sensitivity}.

We also provided a comparison of the sensitivity that can be achieved with an interferometric sequence that involves multiple closed diamonds in spacetime, the so-called resonant mode. As Fig.~\ref{fig:resonant sensitivity} showed, utilising a multiple diamond configuration did not ultimately result in an improvement in sensitivity over the broadband sequence.

Although our focus has been on the signal induced by ultra-light scalar dark matter, we end by commenting on the potential impact of our study for gravitational wave searches with atom gradiometers.
The~$\Delta r/L$ correction will also arise in a careful calculation of the signal induced by a gravitational wave travelling through an atom gradiometer.
This correction can be particularly relevant in long-baseline, single-photon, {\it multi}-gradiometer configurations (i.e.\ when more than two atom interferometers are located within a single baseline and are referenced by the same laser), which have been envisaged to improve GGN characterisation. A combined analysis of the signal-phases recorded by different pairs of interferometers would have to take into account the different gradiometer lengths to ensure accurate predictions.

\section*{Acknowledgements}

We are grateful to members of the AION Collaboration for many fruitful discussions and
we thank Ankit Beniwal, Elliot Bentine, Thomas Edwards, John Ellis, Tiffany Harte and Thomas Hird for comments on the manuscript.
L.B.\, D.B.\ and C.M.\ acknowledge support from the Science and Technology Facilities Council (STFC) Grant No.\ ST/T00679X/1. In addition, L.B.\ is a recipient of an STFC quota studentship and C.M.\ is supported by the STFC Grant No.\ ST/N004663/1.
D.B.\ acknowledges support from the Fundaci\'on Jesus Serra and the Instituto de Astrof\'isica de Canarias under the Visiting Researcher Programme agreed between both institutions. IFAE is partially funded by the CERCA program of the Generalitat~de~Catalunya.
DB is supported by a `Ayuda Beatriz Galindo Senior' from the Spanish `Ministerio de Universidades', grant BG20/00228. The research  of DB leading to these results has received funding from the Spanish Ministry of Science and Innovation (PID2020-115845GB-I00/AEI/10.13039/501100011033).\\

\appendix

\section{Scalar ULDM phase calculation}
\label{A2}

As a result of the linear interactions introduced in Eq.~\eqref{DM-SM lin couplings}, \textit{all} electronic energy levels in an atom receive an oscillating ULDM-induced correction proportional to the DM field. Therefore, the energies of the ground and excited states in the two level system considered in this work, namely $\mathrm{5s^2\,^1S_0}$ and $\mathrm{5s 5p\,^3P_0} $ in $^{87}\mathrm{Sr}$ respectively, are functions of space and time and oscillate at a frequency largely set by the DM mass. Neglecting the spatial variation of the field, we define the ULDM-induced time variation of the $\mathrm{5s^2\,^1S_0}$ and $\mathrm{5s 5p\,^3P_0}$ state energies as $\Delta \omega_g (t)$ and $\Delta \omega_e (t)$, respectively, and they satisfy the relation $\Delta \omega_e (t) = \Delta \omega_g (t) + \Delta \omega_A (t)$, where $ \Delta\omega_A (t)$ is the oscillating transition energy correction defined in Eq.~\eqref{eq: oscillating energy 2}. 

The ULDM-induced oscillations of fundamental constants would be observable in the evolution of the \textit{internal} degrees of freedom of the atom between atom-light interaction points. Neglecting the spatial evolution of the DM field, the ULDM-induced phase contribution between times $t_1$ and $t_2$ along \textit{any} path is
\begin{equation} 
\Phi_{t_1, (g,e)}^{t_2} \equiv \int_{t_1}^{t_2} {\Delta \omega_{(g,e)}(t) dt} \,,
\label{eq: ULDM path diff}
\end{equation}
where the subscripts $g$ and $e$ label the ground and excited state energies, respectively.

In the limit that the separation between the upper and lower arms of the interferometer at the end of the sequence is small, to leading order the interferometer sequence can be modelled as a closed loop (see Fig.~\ref{fig:space_time}). Combining Eq.~\eqref{eq: ULDM path diff} with the definitions of $\Delta \omega_{g,e}$, and computing the difference between the sum of all ULDM-induced phase contributions along each interferometer arm, the total ULDM-induced phase difference of the $p^{\mathrm{th}}$ interferometer relative to the atom's ground state takes the form
\begin{equation}
    \Phi_{p,\mathrm{tot}} \approx \Phi_p + \oint_C \Delta \omega_{g}(t) \, dt = \Phi_p\;.
\end{equation}
In the limit that the separation phase is negligible, the integral is a closed-loop integral over $\Delta \omega_g(t)$, which is identically zero.
Here, $\Phi_p$ is the quantity defined in Eq.~\eqref{full signal AI}, which is the total ULDM-induced phase difference of the $p^{\mathrm{th}}$ single-photon atom interferometer accumulated in the propagation phase of the excited state ($\mathrm{5s 5p\,^3P_0}$) {\it relative} to the ground state ($\mathrm{5s^2\,^1S_0}$). 

Thus, in the limit that the separation phase is negligible, the total ULDM-induced phase difference of a lone single-photon atom interferometer only depends on the phase accumulated by the excited state relative to the ground state of a two-level system. 

\begin{widetext}

\section{Signal calculation for compact atom gradiometers}
\label{A1}

In the range of frequencies probed by atom gradiometers, $10^{-1}\,\mathrm{Hz} \lesssim \omega_\phi \lesssim 10^5\, \mathrm{Hz}$, the spatial dependence of the ULDM field is highly subdominant and can be safely neglected, i.e.\ $\mathbf{k}_\phi\cdot\mathbf{r} \ll 1$. 
Therefore, the expression for the gradiometer signal, Eq.~\eqref{full signal}, contains pairwise contributions that are identical up to time integration limits, where the difference depends on the position of the atomic wave-packets during atom-laser interactions. 
Explicitly, using Eq.~\eqref{full signal AI} and grouping terms, Eq.~\eqref{full signal} can be expressed~as 
\begin{equation}
\begin{aligned}
\Phi_{\mathrm{s}} = & \sum_{m=1}^{n/2} \left [ \Phi_{T+(2m-n-1)L+r_{1,l}}^{T+(2m-n+1)L-r_{1,l}}-\Phi_{T+(2m-n-1)L+r_{2,l}}^{T+(2m-n+1)L-r_{2,l}} \right ]  
 + \sum_{m=1}^{n/2} \left [  \Phi_{2T-(n-2m)L-r_{1,f}}^{2T-(n-2m)L+r_{1,f}} - \Phi_{2T-(n-2m)L-r_{2,f}}^{2T-(n-2m)L+r_{2,f}} \right ] \\
 & - \sum_{m=1}^{n/2} \left [ \Phi_{T+(2m-1)L-r_{1,u}}^{T+(2m-1)L+r_{1,u}} - \Phi_{T+(2m-1)L-r_{2,u}}^{T+(2m-1)L+r_{2,u}} \right ] - \sum_{m=1}^{n/2} \left [ \Phi_{(2m-2)L+r_{1,i}}^{2mL-r_{1,i}}- \Phi_{(2m-2)L+r_{2,i}}^{2mL-r_{2,i}} \right ]\;.
\end{aligned}
\label{full signal step 1}
\end{equation}
Each of the four terms in square brackets can be manipulated into a simpler form by performing the integrals in the ULDM-induced phase contribution as defined in Eq.~\eqref{integral1}, using trigonometric identities, and summing over the LMT kicks. We will show the manipulations explicitly only for the final square bracket term as it is straightforward to adapt the argument for the other terms. 

We begin by performing the integral in the general expression for the ULDM-induced phase contribution, $\Phi_{t_1}^{t_2}$, defined in Eq.~\eqref{integral1}, and simplifying with a sum-to-product trigonometric identity: 
\begin{align}
    \Phi_{t_1}^{t_2}&=\overline{\Delta \omega_A}  \int_{t_1}^{t_2}  \cos (\omega_{\phi} t + \theta)dt\\
    &= \label{eq:Phit2t1} 2\frac{ \overline{\Delta \omega_A}}{\omega_{\phi}} \sin\left(\frac{\omega_{\phi}(t_2-t_1)}{2}\right) \cos\left(\frac{\omega_{\phi}(t_1+t_2)}{2} + \theta\right)\;.
\end{align}
This implies that
\begin{align}
&\sum_{m=1}^{n/2} \left [ \Phi_{(2m-2)L+r_{1,i}}^{2mL-r_{1,i}}- \Phi_{(2m-2)L+r_{2,i}}^{2mL-r_{2,i}} \right ] \\
&=  2\frac{ \overline{\Delta \omega_A}}{\omega_{\phi}} \Big[ \sin\bigl(\omega_{\phi}L- \omega_{\phi}r_{1,i} \bigr) -\sin\bigl(\omega_{\phi} L-\omega_{\phi} r_{2,i} \bigr)  \Big] \sum_{m=1}^{n/2} \cos\bigl(  \omega_{\phi}L (2m-1) + \theta\bigr)\\
&=\label{eq:appB_derive1}2\frac{ \overline{\Delta \omega_A}}{\omega_{\phi}} \left[\frac{ \sin\bigl(\omega_{\phi} L-\omega_{\phi} r_{1,i} \bigr) -\sin\bigl(\omega_{\phi} L-\omega_{\phi} r_{2,i} \bigr)  }{\sin\left( \omega_{\phi}L\right)}\right] \sin\left( \frac{n \omega_{\phi}L}{2}\right) \cos\left(\frac{n \omega_{\phi}L}{2} +\theta \right) \;,
\end{align}
where to reach the final line, we have used a modified form of Lagrange's identity:
\begin{equation}\label{eq:Lagrange}
    \sum_{m=1}^N \cos( m A + B) = \frac{\sin(AN/2)}{\sin(A/2)} \cos\left(\frac{A+2 B +A N}{2} \right) \;.
\end{equation}
We recognise that Eq.~\eqref{eq:appB_derive1} contains the term
\begin{equation}
    \Phi^{nL}_{0} = 2\frac{ \overline{\Delta \omega_A}}{\omega_{\phi}} \sin\left( \frac{n \omega_{\phi}L}{2}\right) \cos\left(\frac{n \omega_{\phi}L}{2} +\theta \right)\;,
\end{equation}
which allows us to write
\begin{equation}
    \sum_{m=1}^{n/2} \left [ \Phi_{(2m-2)L+r_{1,i}}^{2mL-r_{1,i}}- \Phi_{(2m-2)L+r_{2,i}}^{2mL-r_{2,i}} \right ] = \Phi^{nL}_{0} \left[\frac{ \sin\bigl(\omega_{\phi}L-\omega_{\phi} r_{1,i} \bigr) -\sin\bigl(\omega_{\phi} L-\omega_{\phi}r_{2,i} \bigr)  }{\sin\left( \omega_{\phi}L\right)}\right]\;.
\end{equation}

Similar manipulations can be used on the other terms to arrive at the expression
\begin{equation}
\begin{aligned}
\Phi_{\mathrm{s}} &=  \Phi^{T+L}_{T-(n-1)L}\left [ \frac{ \sin{(\omega_\phi L - \omega_\phi r_{1,l})}-\sin{(\omega_\phi L - \omega_\phi r_{2,l})} }{\sin\left(\omega_\phi L \right)}\right] + \Phi^{2T+L}_{2T-(n-1)L}\left [ \frac{ \sin\left(\omega_\phi r_{1,f}\right)-\sin\left(\omega_\phi r_{2,f}\right) }{\sin\left(\omega_\phi L \right)}\right]  \\
 &\qquad - \Phi^{T+nL}_{T}\left [ \frac{ \sin\left(\omega_\phi r_{1,u}\right)-\sin\left(\omega_ \phi r_{2,u}\right) }{\sin\left(\omega_\phi L \right)}\right] - \Phi^{nL}_{0} \left [ \frac{ \sin{(\omega_\phi L - \omega_\phi r_{1,i})}-\sin{(\omega_\phi L - \omega_\phi r_{2,i})} }{\sin\left(\omega_\phi L \right)}\right] \, .
\end{aligned}
\label{full signal step 2}
\end{equation}

So far, we have made no assumption about the position vectors.
However, if two atom interferometers have the same experimental parameters, as we have been assuming in this paper, then
some additional simplifications can be made so that $\Phi_{\mathrm{s}}$ only depends on the distance between the atom interferometers, $\Delta r$, which was previously defined in Eq.~\eqref{eq:deltar_defintion}, and the mean distance $\bar{r}_a=(r_{1,a}+r_{2,a})/2$, where $a$ labels the location of the atom during the sequence. In this case, we can use the sum-to-product trigonometric identities to write
\begin{equation}\label{full signal step 3}
\begin{split}
    \Phi_{\mathrm{s}} = \frac{2 \sin\left( \omega_{\phi} \Delta r/2\right)}{\sin\left( \omega_{\phi} L\right)} &\Big[ \Phi^{T+L}_{T-(n-1)L} \cos\left( \omega_{\phi}L-\omega_{\phi}\bar{r}_l \right) + \Phi^{2T+L}_{2T-(n-1)L} \cos\left( \omega_{\phi}\bar{r}_f\right) \\ 
    &\qquad -\Phi^{T+nL}_{T} \cos\left(\omega_{\phi}\bar{r}_u \right) - \Phi^{nL}_{0} \cos\left( \omega_{\phi}L-\omega_{\phi}\bar{r}_i\right) \Big]\;.
    \end{split}
\end{equation}
As discussed earlier in the paper, since $\omega_\phi L \ll 1$ and $r_{1,i}, r_{1,u},r_{1,l},r_{1,f},r_{2,i}, r_{2,u},r_{2,l},r_{2,f} < L$, the sine and cosine terms in Eq.~\eqref{full signal step 3} can be expanded to give
\begin{equation}
\Phi_{\mathrm{s}} = \frac{\Delta r}{L}\Big \{ \left[ \Phi^{T+L}_{T-(n-1)L} - \Phi^{nL}_{0} \right] - \left[ \Phi^{2T+L}_{2T-(n-1)L} - \Phi^{T+nL}_{T} \right] \Big \} +\mathcal{O}\Big( \left ( \omega_{\phi} L \right)^2 \Big) \,. 
\label{full ULDM phase}
\end{equation}
 In AION-10, where $L\approx10$~m, the next-to-leading term in the expansion at $\omega_{\phi}\sim 1~\mathrm{Hz}$ is suppressed by a factor of $(\omega_{\phi}L)^2 \sim 10^{-15}$; hence, to very high precision the ULDM-induced differential phase is given by the leading-order term in $\Delta r/L$ in Eq.~\eqref{full ULDM phase}, which is the expression that is given in Eq.~\eqref{approximate signal amplitude}.

Finally, we can perform some further straightforward trigonometric
manipulations on Eq.~\eqref{full ULDM phase} to arrive at a compact expression for $\Phi_{\mathrm{s}}$ and 
therefore show the origin of Eq.~\eqref{signal amplitude}. 
Starting with Eq.~\eqref{eq:Phit2t1} and by using cosine sum-to-product identities, we find
\begin{align}
    \left[ \Phi^{T+L}_{T-(n-1)L} - \Phi^{nL}_{0} \right] &= -4 \frac{ \overline{\Delta \omega_A}}{\omega_{\phi}} \sin\left[\frac{\omega_{\phi}n L}{2}\right] \sin \left[\frac{\omega_{\phi}(T-(n-1)L)}{2} \right]\sin \left[\frac{\omega_{\phi}(T+L)}{2} +\theta \right]\\
    \left[ \Phi^{2T+L}_{2T-(n-1)L} - \Phi^{T+nL}_{T} \right] & = -4 \frac{ \overline{\Delta \omega_A}}{\omega_{\phi}} \sin\left[\frac{\omega_{\phi}n L}{2}\right] \sin \left[\frac{\omega_{\phi}(T-(n-1)L)}{2} \right]\sin \left[\frac{\omega_{\phi}(3 T+L)}{2} +\theta \right]\;.
\end{align}
Substituting these expressions into the leading order term in Eq.~\eqref{full ULDM phase}, together with a final application of a sine sum-to-product identity, we arrive at
\begin{equation}
\Phi_{\mathrm{s}} = 8  \frac{ \overline{\Delta \omega_A}}{\omega_{\phi}} \frac{\Delta r}{L} \sin\left[\frac{\omega_{\phi}n L}{2}\right] \sin \left[\frac{\omega_{\phi}(T-(n-1)L)}{2} \right]  \sin\left[\frac{\omega_{\phi} T}{2}\right] \cos\left[ \omega_{\phi}T + \frac{\omega_{\phi}L}{2} + \theta \right]\;.
\label{eq:full signal with phase}
\end{equation}
The last term in Eq.~\eqref{eq:full signal with phase} contains information about the phase of the DM field and will not be observable over the measurement campaign. In fact, because we assume that we are working in the regime where the coherence time of the signal over the relevant mass range is shorter than the projected interrogation time, all phase information will be lost. The estimator of the signal's power spectral density (PSD) would then be the appropriate analysis to extract information from the time-dependent signal collected over the course of a measurement campaign~\cite{VanderPlas_2018,Foster:2017hbq}. In particular, this quantity contains information on the signal's frequency spread and is proportional to the amplitude squared of the analysed signal at $\omega_\phi$~\cite{Foster:2017hbq}. 
The amplitude of the signal (i.e. Eq.~\eqref{eq:full signal with phase}) then takes the form as presented in Eq.~\eqref{signal amplitude}, explicitly
\begin{align}
\overline{\Phi}_{\rm{s}} &= \sqrt{\frac{1}{\pi}\int_{0}^{2\pi}\Phi_s^2 \, d\theta} \\ 
&=  8 \frac{\overline{\Delta \omega_{A}}}{m_{\phi}}\frac{\Delta r}{L} \Bigg |\sin \left[\frac{m_{\phi}nL}{2}\right] \sin \left[\frac{m_{\phi} T}{2}\right] \sin \left[\frac{m_{\phi}(T-(n-1)L)}{2}\right]  \Bigg |\label{eq:broadband}\;,
\end{align}
where we have neglected sub-leading kinetic corrections to the angular frequency by making the approximation $\omega_\phi \approx m_\phi$. 

\section{Signal calculation for the case of multiple diamonds }
\label{A_diamonds}

The ULDM-induced phase calculation presented in Appendix~\ref{A1} applies to the `single-diamond' configuration shown in Fig.~\ref{fig:space_time}. However, we also considered in section~\ref{sec:resonant} the resonant mode search that employs a sequence with~$Q$ copies of the single-diamond configuration (see the left inset in Fig.~\ref{fig:resonant sensitivity} for an example with $Q=2$). In this appendix, we derive the form of the ULDM-induced phase for a sequence with $Q$-diamonds, where each diamond contains $n$ LMT kicks. This implies that the full sequence has a total of $[2Q(2n-1)+1]$ laser pulses and we use the labelling convention that the final $\pi/2$-pulse is emitted after a time $2QT$.

Repeating the steps in Appendix~\ref{A1} but for a sequence that starts at time $(2q-2)T$, it is straightforward to show that the induced phase for the $q^{\mathrm{th}}$-diamond, which lasts for a duration of approximately $2T$, is
\begin{align}
\Phi_{\mathrm{s}}^q &= \frac{\Delta r}{L}\Big \{ \left[ \Phi^{(2q-1)T+L}_{(2q-1)T-(n-1)L} -    \Phi^{(2q-2)T+nL}_{(2q-2)T}  \right] - \left[ \Phi^{2qT+L}_{2qT-(n-1)L} - \Phi^{(2q-1)T+nL}_{(2q-1)T} \right] \Big \} \label{qthULDMphase} \\
& = 8  \frac{ \overline{\Delta \omega_A}}{\omega_{\phi}} \frac{\Delta r}{L} \sin\left[\frac{\omega_{\phi}n L}{2}\right] \sin \left[\frac{\omega_{\phi}(T-(n-1)L)}{2} \right]  \sin\left[\frac{\omega_{\phi} T}{2}\right] \cos\left[ (2q-1) \omega_{\phi}T + \frac{\omega_{\phi}L}{2} + \theta \right]\;.
\end{align}
Here, we have ignored the negligible corrections $\mathcal{O}\left( \left ( \omega_{\phi} L \right)^2 \right)$. The full induced phase is obtained by summing the contribution from all diamonds, which, with the help of Eq.~\eqref{eq:Lagrange}, gives
\begin{equation}
\sum_{q=1}^{Q}\Phi_{\mathrm{s}}^q  = 8  \frac{ \overline{\Delta \omega_A}}{\omega_{\phi}} \frac{\Delta r}{L} \sin\left[\frac{\omega_{\phi}n L}{2}\right] \sin\left[\frac{\omega_{\phi}T}{2}\right] \sin \left[\frac{\omega_{\phi}(T-(n-1)L)}{2} \right] \frac{\sin\left[Q \omega_{\phi} T\right]}{\sin\left[\omega_{\phi} T\right]}   \cos\left[ Q \omega_{\phi}T + \frac{\omega_{\phi}L}{2} + \theta \right] \;.
\end{equation}
From this expression and when the sub-leading kinetic corrections are ignored, we obtain the resonant mode signal amplitude
\begin{equation}
\overline{\Phi}^{\,Q}_{\rm{s}}  = 8  \frac{ \overline{\Delta \omega_A}}{m_{\phi}} \frac{\Delta r}{L} \Bigg | \sin\left[\frac{m_{\phi}n L}{2}\right] \sin\left[\frac{m_{\phi}T}{2}\right] \sin \left[\frac{m_{\phi}(T-(n-1)L)}{2} \right] \frac{\sin\left[Q m_{\phi} T\right]}{\sin\left[m_{\phi} T\right]} \Bigg |,
\end{equation}
which is the amplitude of the signal presented in Eq.~\eqref{eq:resonant mode amp}. This result differs from Eq.~\eqref{eq:broadband} by the factor  $\sin\left[Q m_{\phi} T\right]/\sin\left[m_{\phi} T\right]$.

\end{widetext}

\bibliographystyle{bibi}
\bibliography{ref}

\end{document}